\documentstyle[12pt]{article}

\newtheorem{Theorem}{Theorem}
\newtheorem{Lemma}{Lemma}

\def\bbbc{{\mathchoice {\setbox0=\hbox{$\displaystyle\rm C$}\hbox{\hbox
to0pt{\kern0.4\wd0\vrule height0.9\ht0\hss}\box0}}
{\setbox0=\hbox{$\textstyle\rm C$}\hbox{\hbox
to0pt{\kern0.4\wd0\vrule height0.9\ht0\hss}\box0}}
{\setbox0=\hbox{$\scriptstyle\rm C$}\hbox{\hbox
to0pt{\kern0.4\wd0\vrule height0.9\ht0\hss}\box0}}
{\setbox0=\hbox{$\scriptscriptstyle\rm C$}\hbox{\hbox
to0pt{\kern0.4\wd0\vrule height0.9\ht0\hss}\box0}}}}

\begin{document}
\date{}

\title{\bf A note on quantum black-box complexity 
of almost all Boolean functions}

\author{Andris Ambainis\thanks{Address: Computer Science Division, 
       University of California, Berkeley, CA 94720-1776,
       e-mail: {\tt ambainis@cs.berkeley.edu}.
       Supported by Berkeley Fellowship for Graduate Studies.}\\
       UC Berkeley}
       
\maketitle

\begin{abstract}
We show that, for almost all $N$-variable Boolean functions $f$,
$N/4-O(\sqrt{N}\log N)$ queries are required to compute $f$
in quantum black-box model with bounded error.
\end{abstract}

\section{Introduction}

In the black-box computation model, we assume that the input 
are given by a black box that, given an index $i$, returns the $i^{\rm th}$
bit of the input. Several efficient quantum algorithms can be 
considered in this framework, including Grover's algorithm\cite{Grover95} 
and many its variants.  

Beals, Buhrman et.al.~\cite{Beals98} proved that almost all $N$-variable 
Boolean functions require $\Omega(N)$ queries in this model if 
the computation has to be exact (i.e., no error is allowed).
We extend their result to computation with bounded error. 

In this case, a non-trivial speedup is possible.
Namely, van Dam\cite{Dam98} showed that all $N$ input bits can be
recovered with just $N/2+o(N)$ queries and arbitrarily small probability
of error. This allows to compute any function with just $N/2+o(N)$ queries.
This bound is known to be tight (up to $o(N)$ term) 
for the parity function\cite{Beals98,Sipser98}
but not for other functions.

In this paper, we show that almost all Boolean functions require 
$N/4-O(\sqrt{N}\log N)$ queries in the quantum black-box model.
This matches van Dam's result up to a constant factor
($N/4$ compared to $N/2$).

\section{Quantum black-box model}

We consider computing a Boolean function 
$f(x_1, \ldots, x_N):\{0, 1\}^N\rightarrow\{0, 1\}$ 
in the quantum black-box model\cite{Beals98}.
In this model, input bits can be accessed by queries to an oracle $X$
and the complexity of $f$ is the number of queries needed to compute $f$.
 
A computation with $T$ queries
is just a sequence of unitary transformations 
\[ U_0\rightarrow O_1\rightarrow U_1\rightarrow O_1\rightarrow\ldots
\rightarrow U_{T-1}\rightarrow O_T\rightarrow U_T\]
on a state space with finitely many basis states.
We shall assume that the set of basis states is $\{0, 1, \ldots, 2^m-1\}$
for some $m$. (Then, $U_0, O_1, \ldots, U_T$ are transformations on $m$ qubits.)

$U_j$'s are arbitrary unitary transformations that do not depend
on $x_1, \ldots, x_N$ and $O_j$ are queries to the oracle.
To define $O_j$, we represent basis states as $|i, b, z\rangle$ where 
$i$ consists of $\lceil \log N\rceil$ bits, $b$ is one bit and 
$z$ consists of all other qubits. Then, $O_j$ maps
$|i, b, z\rangle$ to $|i, b\oplus x_i, z\rangle$.
(I.e., the first $\lceil\log N\rceil$ qubits are interpreted as an index
$i$ for an input bit $x_i$ and this input bit is XORed on 
the next qubit.)

We start with a state $|0\rangle$, apply $U_0$, $O_1$, $\ldots$, $O_T$, 
$U_T$ and measure the rightmost bit of the final state. The network computes
$f$ exactly if, for every $x_1, \ldots, x_N$, 
the result of the measurement always equals $f(x_1, \ldots, x_N)$.
The network computes $f$ with bounded error if, for every $x_1, \ldots, x_N$,
the probability that the result equals  $f(x_1, \ldots, x_N)$ is at
least $2/3$.

For more information about this model, see \cite{Beals98}.

\section{Result}

We are going to prove that almost all $N$-variable functions
$f(x_1, \ldots, x_N)$ require at least 
$T(N)=\frac{N}{4}-2\sqrt{N}\log N$ queries
in the quantum black box model.
First, we state a useful lemma from \cite{Beals98}.

\begin{Lemma}
\cite{Beals98}
Assume we have a computation in the black-box model with $T$ queries.
Then, the probability that the measurement at the end of computation
gives 0 (or 1) is a polynomial $p(x_1, \ldots, x_N)$ of degree at most $2T$.
\end{Lemma}

If a black-box computation computes $f(x_1, \ldots, x_N)$ with a bounded error,
$p(x_1, \ldots, x_N)$ must be in the interval $[2/3, 1]$
if $f(x_1, \ldots, x_N)=1$ and in $[0, 1/3]$ if $f(x_1, \ldots, x_N)=1$.
In this case, we say that $p$ {\em approximates} $f$.
We show that, for almost Boolean functions, there is no polynomial $p$
of degree $2T$ that approximates $f$. 
We start by bounding the coefficients of $p$.

\begin{Lemma}
\label{L1}
If a polynomial $p(x_1, \ldots, x_N)$ approximates a Boolean function 
$f(x_1, \ldots, x_N)$, then 
coefficients of all its $d^{\rm th}$ degree terms are between 
$-2^{Nd+1}$ and $2^{Nd+1}$.
\end{Lemma}

\noindent
{\bf Proof:} 
By induction. 

{\bf Base case:} 
$k=0$. The coefficient is equal to
the value of the polynomial on the all-0 vector, $p(0, \ldots, 0)$.
Hence, it must be between -4/3 and 4/3.

{\bf Inductive case:} 
Let $c$ be the coefficient of $x_{i_1}x_{i_2}\ldots x_{i_d}$.
The value of the polynomial on the assignment with 
$x_{i_1}=\ldots=x_{i_d}=1$ and all other variables equal to 0
is the sum of $c$ and coefficients of all terms that
use part of variables $x_{i_1}, \ldots, x_{i_d}$.
These are terms of degree at most $d-1$. 
Hence, inductive assumption applies to them, 
each of them is at most $2^{N(d-1)+1}$ and
their sum is at most $(2^{d}-1)2^{N(d-1)+1}$.
The sum of this and $c$ should be at most 4/3 by absolute value.
Hence, $|c|$ is at most $(2^d-1)2^{N(d-1)+1}+4/3 < 2^{Nd+1}$.
$\Box$

This implies a bound on the number of polynomials than can be 
approximated. Let $D(N, d)=\sum_{i=0}^d {n \choose i}$.

\begin{Lemma} 
\label{L2}
At most $2^{O(D(N, d)d N^2)}$ functions can be approximated
by polynomials of degree $d$.
\end{Lemma}

\noindent
{\bf Proof:} 
Let $p_1$, $p_2$ be two polynomials. If all coefficients of 
$p_1$ and $p_2$ differ by at most $2^{-N-2}$, the values on any
(0,1)-assignment differ by at most $2^{N}2^{-N-2}=1/4$ 
(since there are at most $2^N$ terms) 
and these two polynomials cannot approximate two different 
Boolean functions. 

By Lemma \ref{L1}, all coefficients of such polynomials are in 
$[-2^{Nd+1},2^{Nd+1}]$. We split this interval into subintervals
of size $2^{-N-2}$. This gives $2^{O(N^2 d)}$ subintervals.
If we choose a subinterval for each coefficient,
there is at most one Boolean function 
approximated by a polynomial with coefficients in these intervals
(because any two such polynomials differ by at most 1/4 and, hence,
cannot approximate different functions).
There are $D(N, d)$ possible terms of degree at most $d$.
Hence, there are at most $(2^{O(N^2 d)})^{D(N, d)}=2^{O(D(N,d)N^2 d)}$ combinations
of intervals.
$\Box$

\begin{Theorem} 
The fraction of Boolean functions that can be 
computed with a bounded error in the quantum black-box model with at
most $T(N)$ queries, for $T(N)=N/4-2\sqrt{N}\log N$, goes to 0, as 
$N\rightarrow\infty$.
\end{Theorem}

\noindent
{\bf Proof:} 
Let $d=2T=N/2-4\sqrt{N}\log N$.
Then, $D(N, d)\leq\frac{2^N}{N^{4}}$.
and $D(N, d) N^2 d\leq D(N, d) N^3 \leq \frac{2^N}{N}$.
Hence, black-box computations with at most
$T(N)=N/4-2\sqrt{N}\log N$ queries can compute
only $2^{\frac{2^N}{N}}=o(2^{2^N})$ functions, 
but there are $2^{2^N}$ different
Boolean functions of $N$ variables.
$\Box$

\end{document}